\documentclass[fleqn,usenatbib]{mnras}

\usepackage{newtxtext,newtxmath}

\usepackage[T1]{fontenc}
\usepackage{ae,aecompl}

\usepackage{graphicx}	
\usepackage{amsmath}	

\title[Space sustainability: a review]{Looking out for a sustainable space}

\author[J. A. Blake]{
\parbox{\textwidth}{
James A. Blake,$^{1,2}$\thanks{E-mail: \href{j.blake.1@warwick.ac.uk}{j.blake.1@warwick.ac.uk}}
}
\\
$^{1}$Department of Physics, University of Warwick, Coventry, CV4 7AL, UK\\
$^{2}$Centre for Space Domain Awareness, University of Warwick, Coventry, CV4 7AL, UK\\
}

\date{Invited review for Astronomy \& Geophysics (RAS Journals)}

\pubyear{2022}

\begin{document}
\label{firstpage}
\pagerange{\pageref{firstpage}--\pageref{lastpage}}
\maketitle

\begin{abstract}
Nearly sixty-five years on from the advent of human activity in space, I chart the evolution of the orbital debris environment and review latest efforts to make operations in the space domain more sustainable.
\end{abstract}

\begin{keywords}
Space Sustainability -- Orbital Debris -- Space Domain Awareness
\end{keywords}



\section{Introduction}
\label{sec:introduction}

October 1957, and the successful launch of Sputnik 1 into Earth orbit, marked the dawn of the Space Age. 
The first of the `fellow travellers' - humanity’s first artificial satellite - orbited for a mere three months before re-entering the Earth’s atmosphere, though its mission paved the way for an era of exploration that has endured to the present day.

For many, a world without satellites would be a difficult one to imagine. 
As a society, we have become reliant on them for a vast array of services and applications. 
With a divine view of large swathes of the Earth’s surface, and the ability to relay signals around its curvature, satellites have enabled the fast transfer of data on a global scale, bypassing the challenges associated with ground-based broadcasting, long-distance wiring, and so on.
Positioning, Navigation and Timing (PNT) satellites have revolutionised transportation by land, air, and sea, while weather satellites enable scientists to monitor and warn of large-scale phenomena as they develop in near real-time. 
Satellites have extended the frontiers of observation: looking outwards, astronomers are able to circumvent the Earth’s atmosphere to look deeper into the cosmos than ever before; looking inwards, patterns and processes that feed into general circulation models can be monitored on a range of timescales, improving our understanding of climate change.

Satellites, and the services they provide, are not going to disappear any time soon. 
That said, threats to satellite safety do exist, and it is important that they be addressed as soon as possible to avoid long-lasting damage to operations in the space domain. 

\section{A short, yet messy history}
\label{sec:a-short-yet-messy-history}

The oldest artificial object in orbit is Vanguard 1, a scientific satellite that launched in 1958. 
Having lost communications in 1964, the satellite has been derelict ever since, residing in an elliptical medium Earth orbit.
Lower-altitude objects, if left to their own devices, will spiral back towards the Earth’s surface over time, owing to drag forces induced by gas molecules in the upper atmosphere. 
This effect acts as a `natural sink' for satellites and orbital debris in the vicinity, though the impact on higher-altitude orbits is negligible~\citep{klinkrad2006space}.
Indeed, Vanguard 1 is expected to remain in orbit for centuries!

\begin{figure*}
	\centering
	\includegraphics[width=0.7\textwidth]{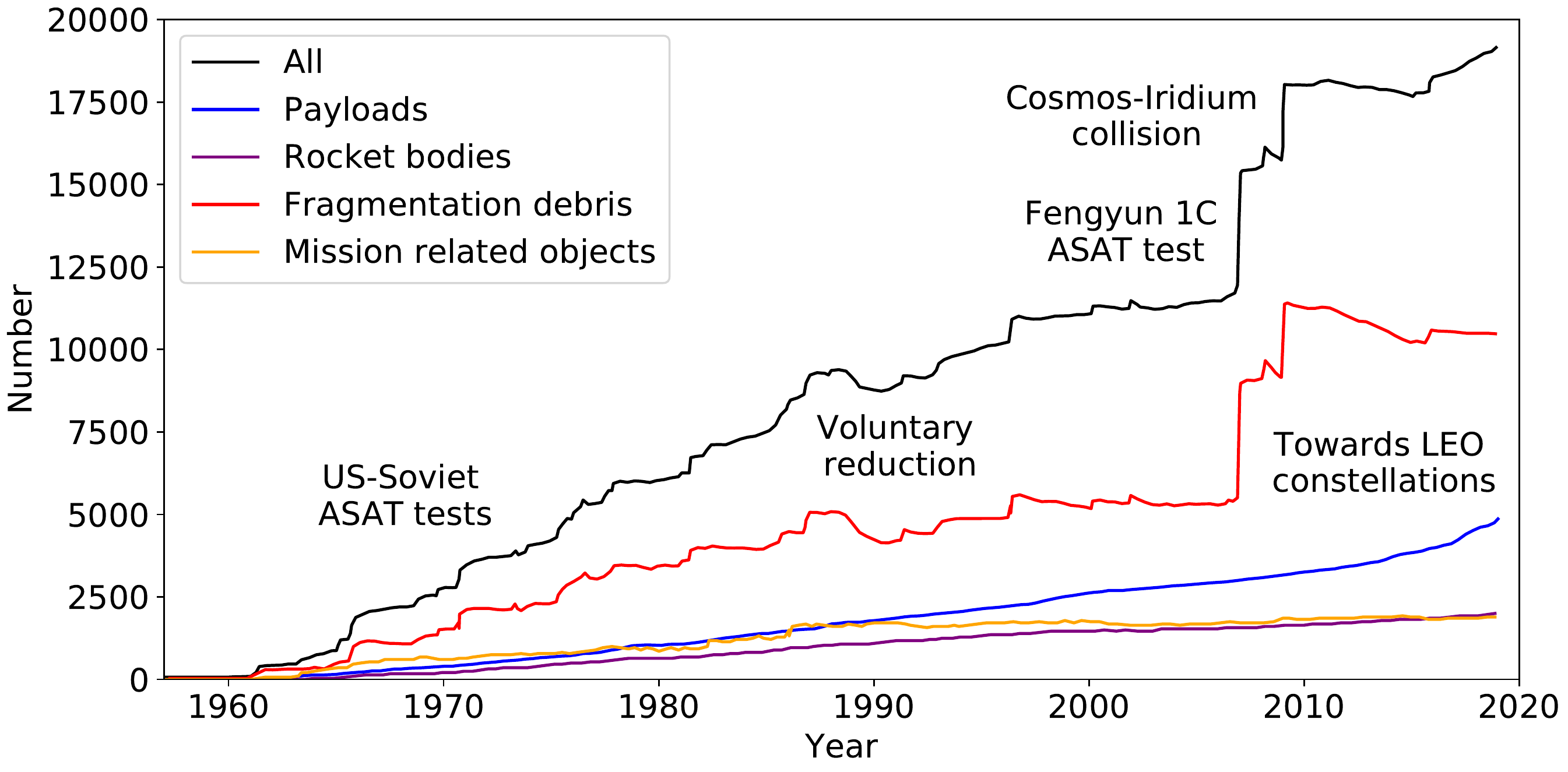}
    \caption{The temporal evolution of the USSPACECOM catalogue for objects larger than 10\,cm in low Earth orbit.
    From~\citet{blake2021optical}, adapted from NASA ODPO.}
    \label{fig:usspacecom-catalogue}
\end{figure*}

The accumulation of defunct satellites and rocket bodies over the years has contributed to a steady rise in the number of orbiting objects tracked and catalogued by surveillance networks. 
In Figure~\ref{fig:usspacecom-catalogue}, the evolution of the most widely used public catalogue is shown, maintained by the US Space Command (USSPACECOM) with observations from the US Space Surveillance Network (SSN), a series of over thirty ground-based radars and optical telescopes across the globe, supported by several satellites in orbit.

\begin{figure*}
	\centering
	\includegraphics[width=0.7\textwidth]{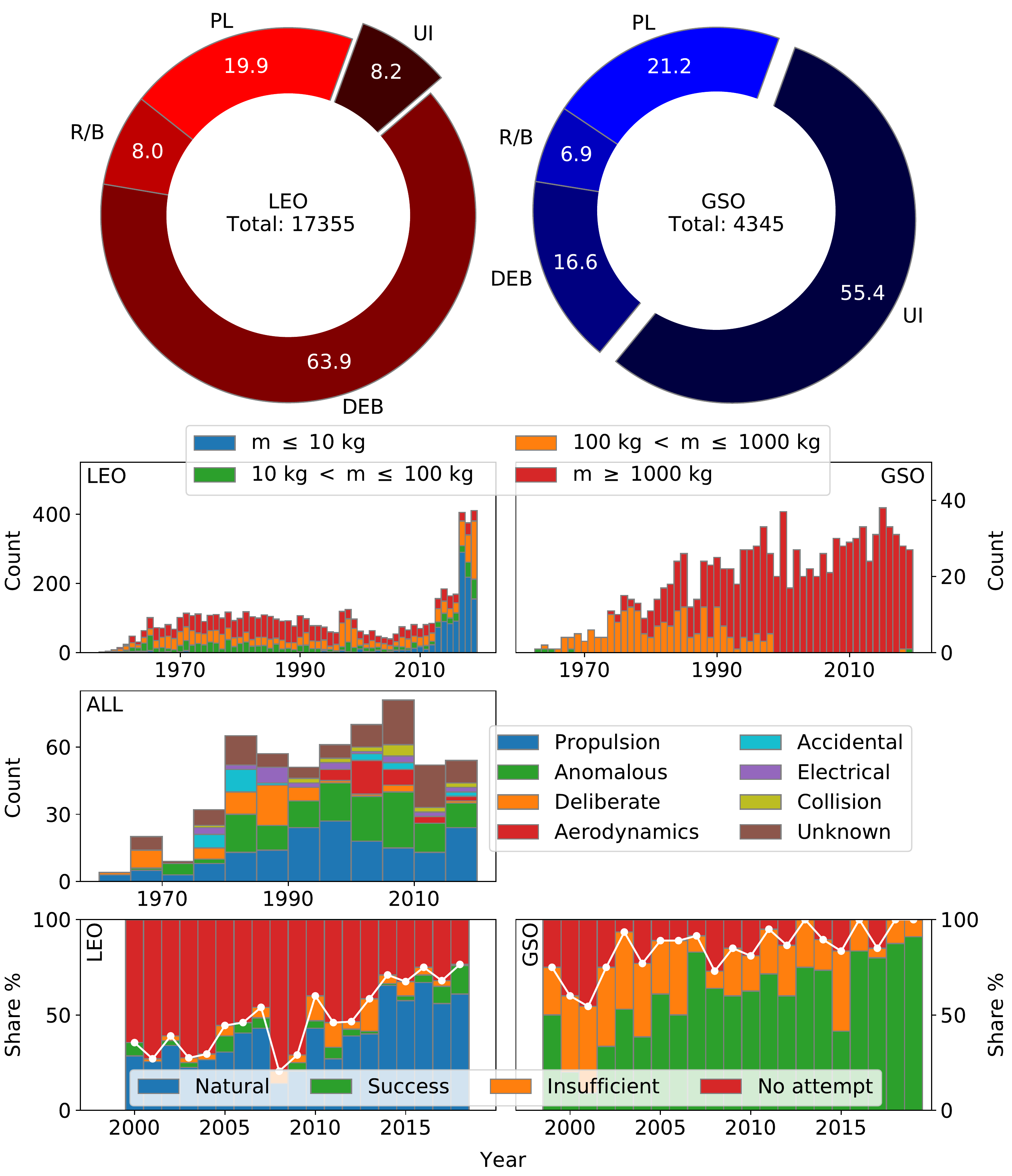}
	\caption{Top row) Number of objects penetrating the LEO (left) and GSO (right) regions by object type: payload (PL), rocket body (R/B), debris (DEB) or unidentified (UI). 
	Second row) Payload launch traffic into the LEO (left) and GSO (right) regions by mass. 
	Third row) Number of fragmentation events by diagnosis. 
	Anomalous fragmentations are those with an unknown cause that leave the parent body largely intact, typically assumed to involve shedding of material or impacts with small/untracked debris.
	Bottom row) Adherence to mitigation guidelines for the LEO (left) and GSO (right) regions, with clearance attempts categorised as naturally compliant, successful, insufficient, or not performed. 
	White lines give the proportion of cases where an attempt was made to adhere to the guidelines. 
	From~\citet{blake2021optical}, information sourced from~\citet{esa2020report}, correct as of the end of 2019.}
	\label{fig:esa-report-stats}
\end{figure*}

While abandoned objects make up a significant proportion of the debris littering the near-Earth environment, by far the most numerous contributions have stemmed from fragmentation events, or `break-ups', in orbit. 
The first of these took place in 1961, when a Thor-Ablestar upper stage exploded shortly after depositing its payload into orbit, producing close to 300 trackable fragments and more than trebling the catalogue at the time~\citep{portree1999orbital}. 
By the end of 2019, the number of confirmed fragmentation events to have taken place in orbit stood at 561~\citep{esa2020report}. 
The most common cause of fragmentations has historically been the explosion of propulsion-related sub-systems (see third row of Figure~\ref{fig:esa-report-stats}) due to thermal stress. 
Explosions can also occur due to overcharging batteries and other electrical failures. 
Operators are now advised to `passivate' their spacecraft at the end of their mission lifetimes, meaning that residual fuel and other reservoirs of stored energy are to be dissipated prior to disposal. 
However, many old derelicts remain in an `unpassivated' state, so explosions will likely continue to supplement the debris population for the foreseeable future.

Perhaps unsurprisingly, not all break-ups have taken place accidentally. 
From the 1960s to the 1980s, the US and Soviet Union frequently demonstrated their anti-satellite (ASAT) capabilities. 
The tens of tests that took place during this period, each generating a few hundred trackable fragments of debris, contributed appreciably to the early growth of the catalogue~\citep{tan1996analysis,portree1999orbital}.
Soviet ASAT tests would typically make use of an interceptor satellite that would detonate when proximate to the target, while the US were the first to deploy a so-called `kinetic kill vehicle' (KKV), namely a homing projectile designed to destroy a satellite via kinetic impact~\citep{tan2019theory}. 

The years of frequent ASAT testing were followed by a two-decade international hiatus, as the major agencies came together to try and tackle the orbital debris problem. 
Fortuitously, the efforts to reduce debris generation coincided with a solar maximum in 1989; the increased solar activity caused the atmospheric envelope to swell, increasing drag and accelerating the re-entry process for debris~\citep{johnson2010orbital}. 
This serendipitous combination resulted in a temporary net decrease in the number of objects being tracked (see Figure~\ref{fig:usspacecom-catalogue}).

It did not take long, however, for the catalogue to resume its growth. 
In 2007, undeterred by the hiatus that had held fast since the end of the Cold War era, China put into practice its own ASAT programme, culminating in the destruction of a defunct meteorological satellite: Fengyun 1C~\citep{kelso2007analysis,liou2009characterization}.
Producing the most trackable fragments of any single event (see the step change in Figure~\ref{fig:usspacecom-catalogue}), the test was widely condemned by the global community. 
Owing to the target’s comparatively high altitude, the break-up produced a cloud of long-lasting debris in an already densely populated region; over half of the fragments catalogued remain in orbit~\citep{lambert2018fengyun}.

Despite an apparent consensus that ASAT tests represent irresponsible and reckless behaviour, legally binding and internationally recognised regulations are still lacking. 
In 2019, India became the fourth nation to demonstrate ASAT capabilities, intercepting the Microsat-R satellite head-on with a KKV. 
The impact is thought to have induced additional on-board explosions, generating several hundred fragments, many of which were propelled into higher-altitude orbits that pose a threat to active satellites~\citep{oltrogge2019characterizing,tan2020posthumous}.
Most recently, in November 2021, the Russian Federation destroyed a derelict spy satellite that launched in 1982. 
Some of the fragments in the resulting debris cloud pass through the orbital altitude of the International Space Station (ISS); indeed, the crew of seven at the time of the test were forced to prepare for emergency evacuation, while surveillance networks assessed the resulting debris field.

As the population densities of orbital regions with useful characteristics have increased over time, so too has the probability of collision in those regions.
Only a handful of accidental collisions between trackable objects have taken place thus far, though `near-miss' events are becoming commonplace. 
In early 2020, for instance, the InfraRed Astronomical Satellite (IRAS) passed within an estimated 11\,m (courtesy: LeoLabs) of the Gravity Gradient Stabilization Experiment (GGSE-4), each decommissioned and orbiting in an uncontrolled state. 
The typical energies involved make this miss distance even more alarming; relative velocities in the low Earth orbit region (LEO, nominally $<$2000\,km altitude) often exceed 10\,kms$^{-1}$, meaning that objects as small as 1\,cm can cause mission-fatal damage. 

The first accidental collision between two satellites took place in 2009, when the active Iridium 33 impacted with a Russian derelict, Cosmos 2251~\citep{kelso2009analysis}. 
The collision was catastrophic, polluting the environment with over 2000 trackable fragments of debris, and likely thousands more that are too small to be detected by current surveillance networks~\citep{wang2010analysis,anz2018history}. 
The Cosmos-Iridium collision took place in an already densely populated altitude band, hosting many of the sun-synchronous satellites carrying out Earth observation, reconnaissance, and weather monitoring. 
The fragments generated by the collision have consequently been responsible for a significant proportion of avoidance manoeuvres undertaken by active satellites since, including those of the ISS itself~\citep{nasa2015quaterly19-1,braun2016operational}.

Several bands of the LEO region are already thought to be unstable, on the cusp of a collisional cascade, whereby collisions generate fragments of debris that go on to seed further collisions, which generate further debris, and so on. 
This scenario is commonly referred to as the `Kessler Syndrome', after Don Kessler, the former NASA scientist who first proposed it~\citep{kessler1978collision}. 
Interestingly, the 2013 film \textit{Gravity} features a Kessler-like cascade seeded by fragments from a spy satellite, itself the target of a Russian ASAT test; it remains to be seen whether the 2021 test will trigger a similar chain of events, though the film should nevertheless serve as a humbling reminder of the potential ramifications of such irresponsible actions.

A short, yet messy history, indeed; change is clearly necessary, so what is currently being done to address the issue?

\begin{figure*}
	\centering
	\includegraphics[width=0.7\textwidth]{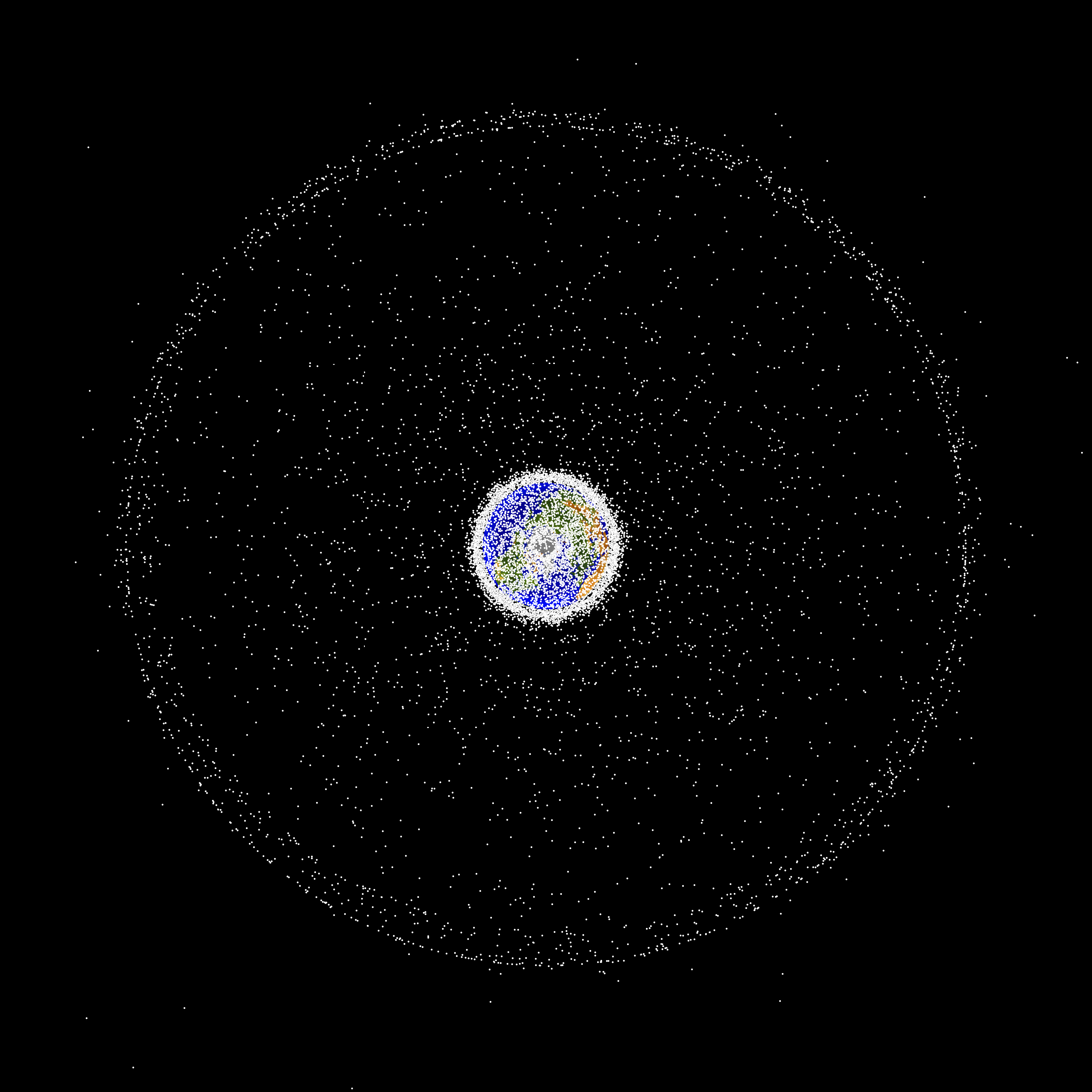}
	\caption{A snapshot of the near-Earth environment, showing the instantaneous location of all objects tracked by the SSN as of January 2019, from a vantage point above the North Pole. 
	Points representing tracked objects (white) are scaled for optimal visibility, not relative to the Earth. 
	From NASA ODPO.}
	\label{fig:usspacecom-artist}
\end{figure*}

\section{Eyes on the sky}
\label{sec:eyes-on-the-sky}

As is true for the US SSN, most space surveillance observations utilise radars or optical telescopes. 
Radars are best suited to monitoring the LEO region, as their sensitivity drops with the fourth power of the range (observer-to-target distance). 
The Royal Air Force Fylingdales radar station in North Yorkshire, for example, serves as a key element of the SSN, tracking objects out to an altitude of roughly 5500\,km. 

High-altitude targets are typically monitored by optical telescopes, with the more favourable inverse square law governing their sensitivities. 
Ground-based Electro-Optical Deep Space Surveillance (GEODSS) is the primary optical component of the SSN, a system of 1\,m telescopes with sites in New Mexico, Hawaii, and the British Indian Ocean Territory. 
Combined with NASA’s new MCAT facility on Ascension Island~\citep{lederer2019mcat}, the sites are tactically chosen to monitor satellites in the geosynchronous (GSO) region, a ring of orbits (see Figure~\ref{fig:usspacecom-artist}) roughly 36000\,km above the Equator, where geostationary satellites reside. 
These are some of the most valuable assets in space, massive buses (see right-hand panel, second row of Figure~\ref{fig:esa-report-stats}) occupying a limited number of longitudinal slots within the geostationary belt.
Satellites in the belt have the unique property of matching the Earth’s rotation, remaining (near-)fixed in an observer’s sky and thus enabling simplified tracking using unidirectional ground-based antennas.

Despite offering the best coverage of global space assets, the USSPACECOM catalogue is far from complete. 
The SSN can reliably track objects down to roughly 5-10\,cm in size at LEO altitudes, and roughly 1\,m when observing higher-altitude regions. 
These cut-offs are a major cause for concern, given that even millimetre-sized debris can prove disruptive in the hypervelocity regime of LEO. 
Indeed, ESA’s Sentinel-1A satellite experienced an anomaly in 2016 that was attributed to a sub-centimetre impactor damaging a solar panel, causing a permanent partial power loss. 
Too small to be registered by the SSN, the characteristics of the colliding fragment were belatedly inferred from images of the resulting crater, taken by an onboard camera~\citep{krag2017sentinel}. 
Even in the GSO region, recent simulations have suggested that objects less than 10\,cm in size could cause mission-fatal damage, despite the comparatively low relative velocities involved~\citep{oltrogge2018comprehensive}.

While satellites and rocket bodies make up over 98\% of the tracked mass in orbit~\citep{esa2020report}, the numerical abundance of small debris remains a critical concern. 
However, very little is known about the nature and extent of this elusive population. 
Bespoke (and thus sporadic) observations using large radars like MIT’s Haystack, JPL’s Goldstone, and others~\citep{stansbery1995characterization,matney1999recent,mehrholz2002detecting,stokely2009debris}, have detected objects in LEO down to a few millimetres in size. 
\textit{In situ} methods have been used to investigate the population of sub-millimetre LEO debris, by examining impact features on spacecraft such as the Long Duration Exposure Facility, Hubble Space Telescope, and Space Shuttle Orbiter~\citep{love1995morphology,graham2001chemistry,christiansen2004space}.
Laboratory-based experiments have attempted to recreate hypervelocity impacts, allowing debris fragments to be studied and characterised~\citep{cowardin2020optical}.

All these sources of statistical information feed into environmental models that operators use to assess the risk posed by orbital debris. 
That said, even the most widely used models, NASA’s ORDEM and ESA’s MASTER, are known to differ significantly in their output~\citep{krisko2015ordem}. 
Most notably, predicted flux densities disagree by more than an order of magnitude in the 1-3\,mm bracket, currently an `observational gap'; too small to be detected by the most sensitive radars, yet too rare to have caused a statistically informative number of craters on surfaces that have been analysed using \textit{in situ} techniques. 
The MASTER model predicts there to be roughly 1 million objects larger than 1\,cm in orbit, while ORDEM predicts a factor of two fewer. 
Both estimates nevertheless serve to highlight the inadequacies of current space surveillance catalogues.

The top row of Figure~\ref{fig:esa-report-stats} shows a breakdown of catalogued objects orbiting in both the LEO and GSO regions. 
A large proportion of objects intersecting the GSO region are `unidentified', their nature and launch origin still unclear. 
There are several reasons for this: historically, the GSO region has received less attention from an orbital debris perspective, and it is likely that numerous anomalous events have taken place without any real-time follow-up; a substantial population of objects with high area-to-mass ratios (HAMRs) exists in the vicinity of the GSO region, having been highly perturbed by solar radiation pressure (SRP) and other poorly modelled forces, rendering it thus far impossible to associate them with a launch~\citep{kelecy2011analysis}; the HAMR objects, likely shreds of blanketing material, typically reside in eccentric orbits, necessitating bespoke observational strategies which have only been employed in the past couple of decades~\citep{fruh2012variation}.

\begin{figure*}
	\centering
	\includegraphics[width=0.7\textwidth]{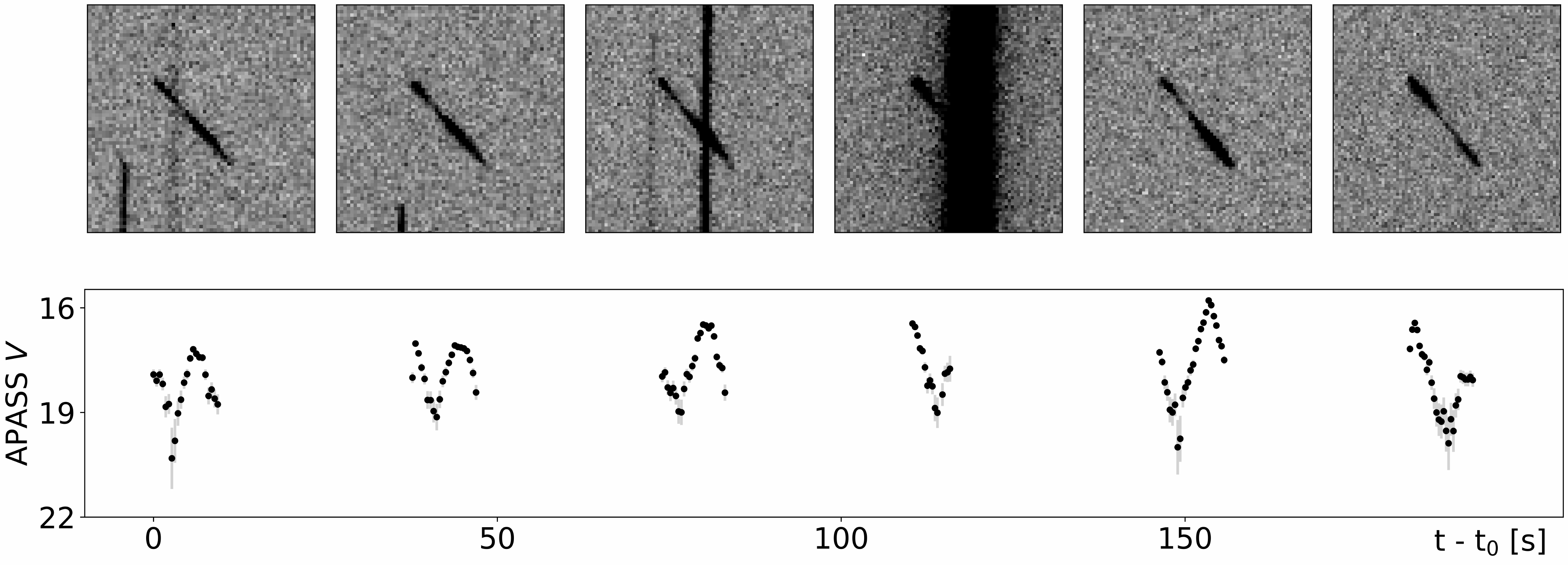}
	\caption{Photometric measurements for a small (20-30\,cm), tumbling fragment of GSO debris, detected by the 2.54\,m Isaac Newton Telescope on La Palma, Canary Islands. 
	From~\citet{blake2021optical}.}
	\label{fig:debriswatch-lightcurve}
\end{figure*}

Recent anomalies and break-ups involving satellites and rocket bodies have added to the relatively mysterious population of small debris in high-altitude orbits~\citep{cunio2017photometric,schildknecht2019esa}. 
In the past couple of decades, several surveys have looked to uncover these faint objects~\citep{schildknecht2007optical,molotov2009faint,blake2020supplementing,blake2021debriswatch}, though cataloguing is rendered very difficult on account of the limited time available with the large aperture telescopes required. 
Indeed, the GSO debris population below roughly 10\,cm in size remains unprobed by any technique. 

Most active satellites conduct regular `station-keeping' manoeuvres to maintain a desired behaviour, as natural perturbations resulting from gravitational asymmetries, atmospheric drag, SRP, and so on, can otherwise cause the spacecraft’s attitude and orbit to evolve~\citep{ziebart2004generalized}.
Observations of uncontrolled debris offer a chance to probe the effects of these perturbations and refine models for predicting attitudinal and orbital states~\citep{papushev2009investigations,kudak2017determining}.
Several groups have acquired multi-colour photometric~\citep{hejduk2012satellite,cardona2016bvri,zigo2019bvri} and time-resolved spectroscopic~\citep{jorgensen2004physical,bedard2014interpretation,vananti2017reflectance} measurements to probe the material properties of objects in orbit, a key consideration when modelling the effects of electro- and geo-magnetic torques~\citep{wetterer2014refining}. 

Many of the small fragments detected in the GSO region have shown photometric signatures of rapid tumbling (see Figure~\ref{fig:debriswatch-lightcurve}), potentially arising from the Yarkovsky-O’Keefe-Radzievskii-Paddack (YORP) effect that is commonly used to explain the rotational dynamics of asteroids and small bodies~\citep{bottke2006yarkovsky,albuja2018yorp}.
Light curves are a useful tool for characterisation, encoding information about an object’s shape and attitude, though it remains a complex task to disentangle desired information from other components~\citep{wetterer2009attitude,hinks2013attitude,fan2020direct}. 
Several groups have conducted bespoke surveys to gather high quality light curves for LEO and GSO objects, with the aim of testing and improving inversion algorithms~\citep{schildknecht2015photometric,chote2019precision,silha2020space}.

Though a key aspect of space domain awareness (SDA), monitoring the mess of near-Earth space cannot solve the problem entirely, especially while the bulk of the dangerous debris population remains invisible and uncatalogued.
Observations alone cannot prevent the explosions of unpassivated derelicts, nor can they thwart the impact of two uncontrolled objects that find themselves on a collision course. 
With no legally binding regulations, the testing of ASAT capabilities can continue to take place unpunished, while operators can opt to launch with little regard for sustainable practices. 
What needs to be done in the future to ensure that satellite operations in the space domain remain sustainable?

\section{Fixes for the future}
\label{sec:fixes-for-the-future}

Policy is one of the major hurdles standing in the way of sustainability in space. 
For the past couple of decades, guidelines developed by the Inter-Agency Space Debris Coordination Committee (IADC), and since adopted by the United Nations Committee on the Peaceful Uses of Outer Space (UNCOPUOS), have formed the basis for standard mitigation practices.
While some space-faring nations have already reflected these guidelines in their national regulatory framework, legally binding regulations are still lacking on the international stage~\citep{weeden2011overview}.

Alongside passivation and collision avoidance, the voluntary guidelines place recommendations on methods for post-mission disposal (PMD). 
The `25-year rule' encourages operators to ensure disposal via atmospheric re-entry within 25 years of mission completion. 
Low-altitude satellites may already be naturally compliant with this, while the deorbit process can be accelerated for non-compliant satellites using propulsion, drag sails, and so on. 
Re-entry is not a viable option for objects in the high-altitude GSO region, so operators will typically attempt to raise decommissioned spacecraft into so-called `graveyard' orbits, clearing the way for future use of the operational zone~\citep{jehn2005situation}. 

Levels of adherence to the debris mitigation guidelines are provided in the bottom row of Figure~\ref{fig:esa-report-stats}. 
Excluding naturally compliant cases, adherence to the 25-year rule in LEO remains concerningly low. 
Recent efforts have looked to formulate a `Space Sustainability Rating' to incentivise operators~\citep{letizia2020space}, though several questions remain unanswered.
With access to space becoming more and more widespread, how can one compare, say, a university-led CubeSat experiment against a constellation operator’s fleet of satellites? 
How does one attribute liability in the event of a collision involving uncontrolled debris? 
How can regulations be drafted in such a way as to ensure an even playing field for emerging nations, when other nations have had ample opportunity to develop sustainable practices through trial and error? 

\begin{figure*}
	\centering
	\includegraphics[width=0.8\textwidth]{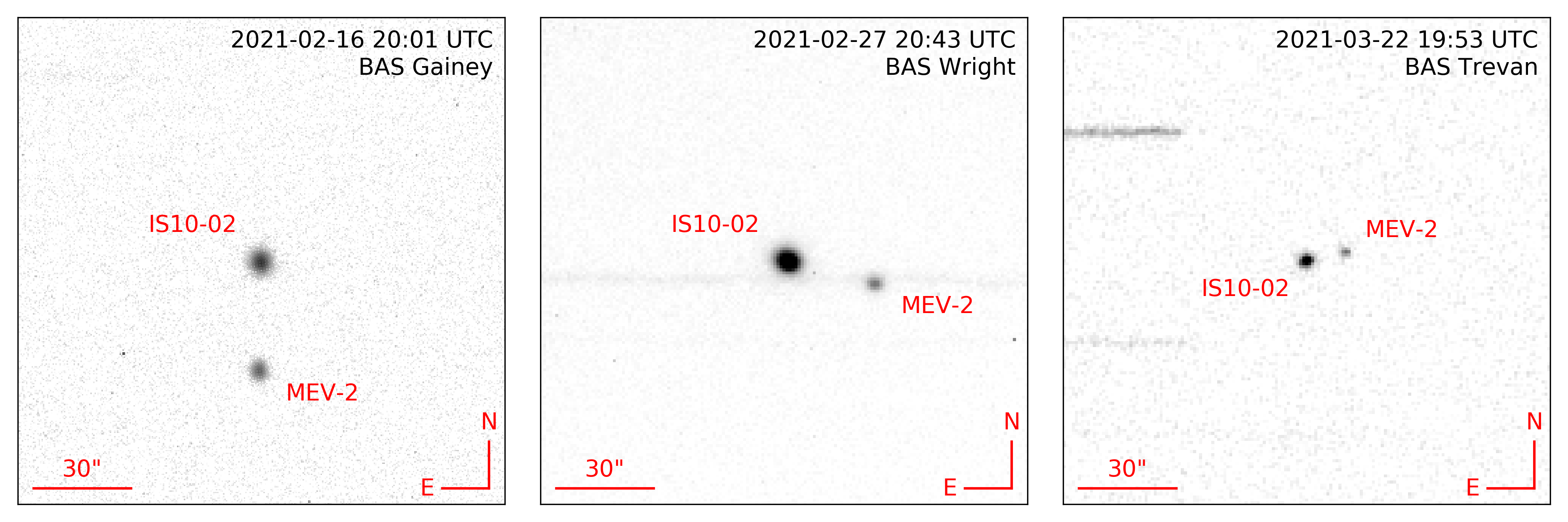}
	\caption{Observations of rendezvous and proximity operations involving the MEV-2 servicer and the Intelsat 10-02 client satellite, acquired by Basingstoke Astronomical Society observers (labelled top-right) during the second \textit{Phantom Echoes} campaign.
	From~\citet{george2021phantom}.}
	\label{fig:mev-2-images}
\end{figure*}

While the legal quagmire surrounding mitigation continues to unfold on the ground, another key focus is on the development and testing of remediation concepts in space. 
Several studies agree that mitigation will need to be supplemented by active debris removal (ADR) to stabilise the LEO environment~\citep{liou2010controlling,bonnal2013active,white2014many}.
Commonly cited as the goal for future ADR activities,~\citet{liou2011active} found the removal of five high priority targets per year to be effective at suppressing a runaway growth in the debris population, when combined with 90\% adherence to PMD guidelines and launch traffic representative of the period 2003-11.
However, with small satellite technologies flourishing in recent years, launch rates consequently soaring, and operators continuing to ignore voluntary standards, it is likely that a higher removal rate will prove necessary.

Large clusters of derelict LEO rocket bodies have been identified as high priority candidates for future ADR missions~\citep{liou2011active,rossi2020short,mcknight2021identifying,maclay2021space}. 
Another key target is ESA’s defunct eight-tonne satellite, Envisat, which resides in a densely populated region and is known to be tumbling~\citep{kucharski2014attitude,pittet2018spin}. 
Studies have explored the challenges of approaching and capturing a tumbling, uncooperative target~\citep{nishida2011strategy,chu2018hybrid}, alongside those associated with the removal of multiple targets per mission~\citep{braun2013active,shen2018optimization}.

A wide variety of mechanisms have been proposed for capturing and removing spacecraft, involving nets~\citep{botta2016simulation,shan2020analysis}, harpoons~\citep{dudziak2015harpoon}, tentacles~\citep{wormnes2013esa,pirat2017mission}, electrodynamic tethers~\citep{kawamoto2006precise,nishida2009space,pardini2009benefits}, drag augmentation~\citep{underwood2019inflatesail}, propulsion~\citep{deluca2013active,olympio2014space}, and laser ablation~\citep{phipps2012removing,soulard2014ican,ebisuzaki2015demonstration}.
Far from exhaustive, this list nevertheless serves to highlight the complexity of the problem, with no one concept offering a `one size fits all' solution. 
More comprehensive reviews of ADR methods have been undertaken by~\citet{shan2016review} and, more recently,~\citet{mark2019review}, though the field continues to evolve rapidly.

Techniques conceptualised for ADR remain largely untested in space. 
The \textit{RemoveDebris} mission, led by Surrey Space Centre, carried out successful captures of planted debris using net and harpoon systems, prior to the partial deployment of a drag sail for re-entry~\citep{forshaw2020active,aglietti2020active}. 
Astroscale’s \textit{ELSA-d} mission launched successfully in March 2021 and aims to demonstrate the capture of a tumbling client in orbit using a magnetic docking system~\citep{forshaw2019elsa}. 
ESA’s \textit{ClearSpace-1} mission will likely be the first to remove an existing object from orbit, currently due to target a Vespa upper stage when it launches to LEO in the mid-2020s~\citep{biesbroek2021clearspace}.

Removal missions targeting higher altitudes remain technologically and financially infeasible, though the GSO region provides an excellent testbed for on-orbit servicing concepts~\citep{xu2011universal,flores2014review,medina2017towards}. 
Northrop Grumman's \textit{Mission Extension Vehicle} (MEV) servicers have successfully rendezvoused and docked with two GSO communications satellites, Intelsat 901 (February 2020) and Intelsat 10-02 (April 2021), with the aim of extending their operational lifetimes. 
The MEV missions have been monitored by a network of Five Eyes sensors as part of the \textit{Phantom Echoes} campaigns~\citep{george2021phantom}.
These have explored the myriad observational challenges associated with rendezvous and proximity operations, from maintaining custody of constantly manoeuvring targets, to resolving the servicer and client satellites as they undergo close approach. 
The \textit{Phantom Echoes} experiment serves as an example of how SDA activities can be coordinated across multiple sectors, combining expertise from government, industry, academia, and even the amateur astronomy community.

\textbf{The Argus Experiment} \textit{Argus} was a citizen science project, bringing together scientists from the Defence Science and Technology Laboratory (UK) and members of the Basingstoke Astronomical Society (BAS) to gather observations of LEO satellites and assess the performance of amateur-grade imaging systems in the context of SDA~\citep{feline2019argus}. 
BAS observers have since gathered a wealth of images for the \textit{Phantom Echoes} experiment, examples of which are provided in Figure~\ref{fig:mev-2-images}, helping to investigate the practical resolution limitations associated with observations of rendezvous and proximity operations with modest equipment.

\section{Looking ahead}
\label{sec:looking-ahead}

What challenges lie ahead? 
Of particular concern are the large constellations of LEO satellites that are under development, offering global coverage and promising ubiquitous access to low latency broadband data services.
Many private companies have already licensed thousands of new satellites, ready to launch into a variety of orbital bands~\citep{muelhaupt2019space,curzi2020large}.
Efforts to simulate the environmental impact of constellations have collectively agreed that adherence to PMD guidelines will be a pivotal factor~\citep{bastidavirgili2016risk,radtke2017interactions,pardini2020environmental}. 
Indeed, the latest update to the US mitigation standard practices states that constellation operators should aim for a 90\% success rate for PMD, though many believe a minimum goal of 99\% will be necessary. 
Aside from PMD, constellations are set to place further strain on the archaic infrastructures of existing surveillance networks; even with responsible management in place, constellation satellites will still be at risk of colliding with uncontrolled debris, thus resulting in more frequent collision avoidance scenarios.

\begin{figure*}
	\centering
	\includegraphics[width=0.45\textwidth]{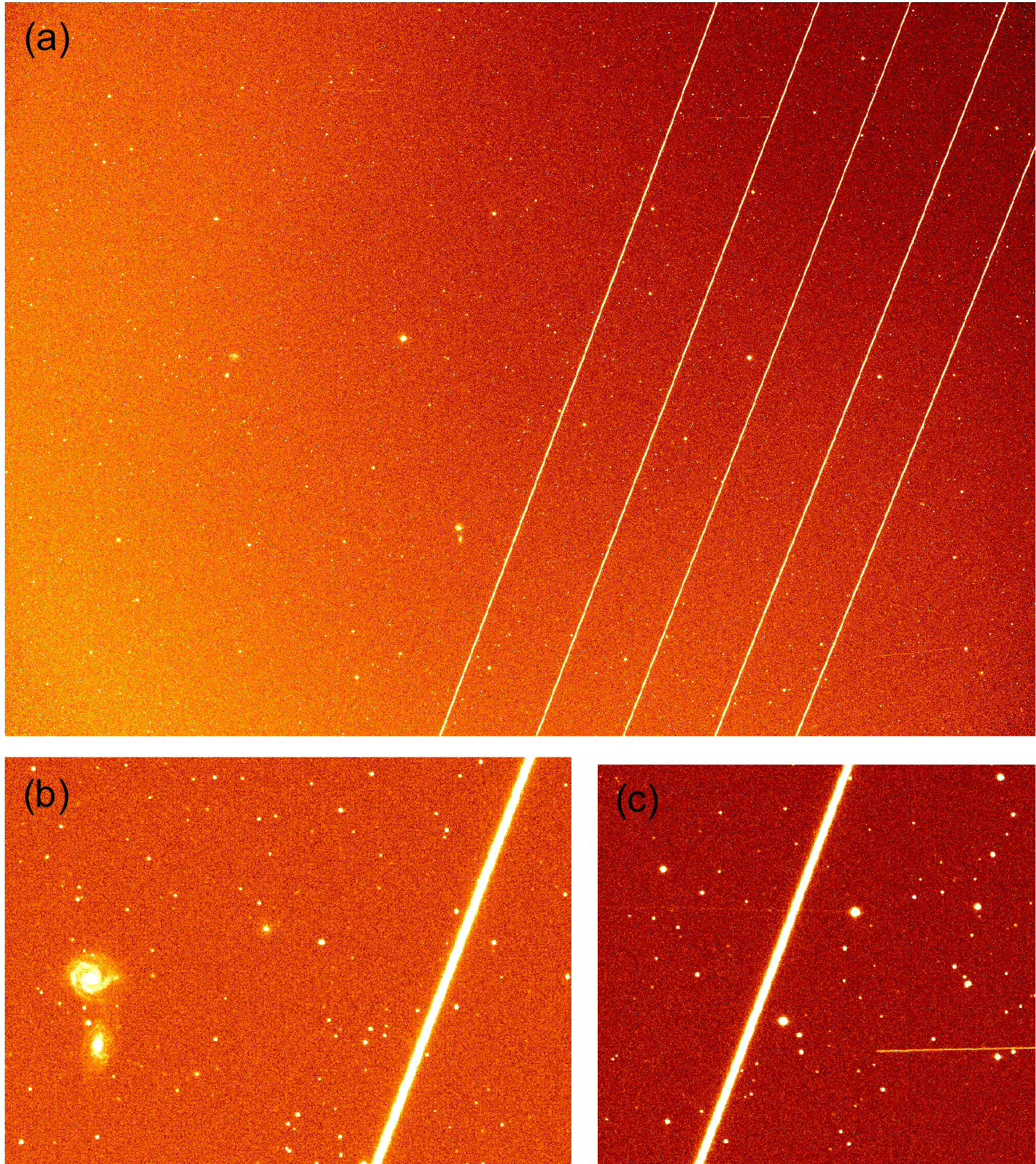}
	\caption{An example of satellite streak contamination in a survey image acquired by the Gravitational wave Optical Transient Observer (GOTO) in La Palma, Canary Islands, a wide-field array focused on detecting the optical counterparts of gravitational wave events~\citep{steeghs2021gravitational}. 
	Close-up views of streaks are provided in (b) and (c), highlighting their superior brightness relative to a galaxy and a higher-altitude satellite, respectively.
	From~\citet{blake2021optical}, courtesy of The GOTO Collaboration.}
	\label{fig:streak-contamination}
\end{figure*}

Numerous studies have highlighted the negative effects that large LEO constellations are likely to have on ground- and space-based astronomical observations across a range of wavelengths~\citep{hainaut2020impact,levchenko2020hopes,mcdowell2020low}. 
Satellite streak contamination in astronomical imaging is by no means a new issue, but the vast numbers and low altitudes involved in maintaining LEO constellations look set to exacerbate the problem, particularly for wide-field systems such as the upcoming Vera C. Rubin Observatory, which will look to study large parts of the sky at any one time, thus resulting in a high probability of field contamination~\citep{massey2020challenge}.
An example of a contaminated wide-field image is provided in Figure~\ref{fig:streak-contamination}. 

While the lowest-altitude constellations are likely to be the brightest, those in higher-altitude bands will perhaps be of greater concern to astronomers; low-altitude satellites will spend much of the night eclipsed in the Earth's shadow, while satellites in the upper bands of the LEO region will remain visible for larger portions of the night. 
This will be the case for nodes of the OneWeb constellation, now part-owned by the UK government. 
OneWeb satellites reside in altitude bands around 1200\,km, to take advantage of a local minimum in the debris population.
\citet{seitzer2020large} has recommended that constellation operators take precautions to keep their satellites faint, and opt for altitude bands below roughly 600\,km, to best combat the issue.

To add to the logistical challenges associated with monitoring a sky that is getting busier every year, surveillance networks may soon be tasked with tracking and cataloguing objects far beyond the `high-altitude' GSO region, namely those in the cislunar domain. 
The expansion of launch traffic into cislunar space in the wake of NASA’s Artemis programme will undoubtedly pose problems for existing SDA architectures~\citep{bolden2020evaluation}: the increased range will result in diminished signal-to-noise, calling for more sensitive instruments; the much larger volume of space in need of monitoring will necessitate a more extensive array of ground- and space-based SDA capabilities; and observations will often be obstructed by the Moon, or eclipsed in shadow, calling for more sophisticated algorithms for object detection and orbital state prediction with sparse or diminished information~\citep{yanagisawa2012shape,virtanen2016streak,hickson2018fast,nir2018optimal,pirovano2020data}. 
It is likely that a variety of astronomical techniques developed for data reduction, classification, fusion, tracking, and association, may prove transferable when applied to many of the upcoming challenges for SDA, from cislunar surveillance to the monitoring of rendezvous and proximity operations for on-orbit servicing and ADR missions.

While orbital debris is not a new problem, it is one that promises to worsen dramatically if the SDA community fails to act. 
The problem is one that affects all operators in space, truly global in nature, and international cooperation will no doubt be integral to the solution, not least in the development of robust regulations for debris mitigation. 
Moreover, the problem warrants a cross-sector, cross-disciplinary approach.
Astronomers are well-placed to help build our understanding of the near-Earth environment, with a wide range of tools and techniques that could complement future SDA missions. 
Surveillance systems will need to handle the strain of monitoring tens of thousands of active satellites, in addition to the hundreds of thousands of fragments that are currently too small to be catalogued, yet sufficiently large to cause significant damage to spacecraft. 
Algorithms for object detection, characterisation, and classification, will need to overcome additional obstacles as rendezvous and proximity operations become commonplace, and orbital traffic starts to spill into the cislunar domain.
Observational data from a plethora of sensors, institutions, nations, will need to be shared and fused (see, for example,
early efforts from the University of Texas at Austin with AstriaGraph, \url{http://astria.tacc.utexas.edu/AstriaGraph/}) to have any hope of preventing collisional cascades from rendering key orbital bands unusable for future generations. 

As new technologies develop, and challenges arise, it is essential that we keep a watchful eye on the sky to protect the satellites that we, as a society, have come to rely on.

\textbf{NAM Space Domain Awareness} The 2021 instalment of the National Astronomy Meeting featured a timely parallel session on SDA, convened by John Zarnecki (Open University) and James Blake (University of Warwick), both active members of the STFC-supported Global Network On Sustainability In Space (GNOSIS).

The session explored the ways in which astronomers and MIST physicists can contribute to enhanced levels of SDA. 
Ian McCrea (STFC) kicked off proceedings with an excellent overview of the importance of monitoring space weather and its effects on atmospheric drag, a key parameter when predicting the positions of satellites. 
Stuart Eves (SJE Space Ltd) moved on to identify the current knowledge and technology gaps that need to be addressed to achieve sufficient levels of SDA, from tracking orbital debris as small as 1\,cm, to determining an object’s origins, intentions, and physical characteristics. 
To round off the session, James Blake presented results from a survey of the geosynchronous region carried out using the Isaac Newton Telescope and a robotic astrograph, on the Canary Island of La Palma. 
Blake concluded with an introduction to GNOSIS, encouraging members of the audience to participate in future workshops.

\textbf{GNOSIS} The STFC-supported Global Network On Sustainability In Space aims to bring together scientists in academia and industry to tackle issues related to the growing problems of orbital debris and space sustainability. 

The network organises `jargon-busting' workshops and conferences for those wishing to learn more about the field. 
It also provides seed funding for proof-of-concept studies and part-funding for studentships in more advanced cases. 
To sign up to monthly newsletters, browse upcoming events, or simply learn more about space sustainability, visit the website: \url{https://gnosisnetwork.org/}.
For regular updates, consider following the network on Twitter at \url{https://twitter.com/gnosis_space}.

\section*{Author}
\label{sec:author}

Dr. James A. Blake is a Research Fellow at the Centre for Space Domain Awareness, University of Warwick, UK, and serves as Secretary to the Steering Board of GNOSIS.
His research focuses on the optical imaging of satellites and orbital debris.

\section*{Acknowledgements}
\label{sec:acknowledgements}

JAB thanks Trevor Gainey, David Wright, Bob Trevan, Paul Chote, the NASA Orbital Debris Programme Office, and the GOTO Collaboration, for use of their figures, alongside Katherine Courtney, Stuart Eves, Will Feline, Simon George, Bob Mann, Don Pollacco and Grant Privett, for their helpful contributions. 
JAB acknowledges support from the Defence Science and Technology Laboratory, UK.



\bibliographystyle{mnras}
\bibliography{references} 




%
%
%
%

\bsp	
\label{lastpage}
\end{document}